# Evidence for a narrow resonance at 1530 MeV/c²
# in the K⁰p-system of the reaction pp → Σ⁺K⁰p
# from the COSY-TOF experiment[*]

## The COSY-TOF Collaboration

M.Abdel-Bary[d], S.Abdel-Samad[b], K.-Th.Brinkmann[b], H.Clement[h], E.Doroshkevich[h], M.Drochner[e], S.Dshemuchadse[b], A.Erhardt[h], W.Eyrich[c,1], D.Filges[d], A.Filippi[g], H.Freiesleben[b], M.Fritsch[a,c], J.Georgi[c], A.Gillitzer[d], D.Hesselbarth[d], R.Jäkel[b], B.Jakob[b], L.Karsch[b], K.Kilian[d], H.Koch[a], J.Kress[h], E.Kuhlmann[b], S.Marcello[f,g], S.Marwinski[d], R.Meier[h], P.Michel[i], K.Möller[i], H.Mörtel[c], H.P.Morsch[d], N.Paul[d], L.Pinna[c], C.Pizzolotto[c], M.Richter[b], E.Roderburg[d], P.Schönmeier[b], W.Schroeder[c], M.Schulte-Wissermann[b], T.Sefzick[d], F.Stinzing[c], G.Y.Sun[b], A.Teufel[c], A.Ucar[d], G.J.Wagner[h], M.Wagner[c,2], A.Wilms[a], P.Wintz[d], St.Wirth[c], P.Wüstner[e]

[a] Institut für Experimentalphysik, Ruhr-Universität Bochum, D-44780 Bochum, Germany
[b] Institut für Kern- und Teilchenphysik, Technische Universität Dresden, D-01062 Dresden, Germany
[c] Physikalisches Institut, Universität Erlangen-Nürnberg, D-91058 Erlangen, Germany
[d] Institut für Kernphysik, Forschungszentrum Jülich, D-52425 Jülich, Germany
[e] Zentral Labor für Elektronik , Forschungszentrum Jülich, D-52425 Jülich, Germany
[f] Dipartimento di Fisica Sperimentale, University of Torino, I-10125 Torino, Italy
[g] INFN, Sezione di Torino, I-10125 Torino, Italy
[h] Physikalisches Institut, Universität Tübingen,, D-72076 Tübingen, Germany
[i] Institut für Kern- und Hadronenphysik, Forschungszentrum Rossendorf, D-01314 Dresden, Germany

_______________________________________________________________________

## Abstract

The hadronic reaction $pp \rightarrow \Sigma^+ K^0 p$ was measured exclusively at a beam momentum of 2.95 GeV/c using the TOF detector at the COSY storage ring. A narrow peak was observed in the invariant mass spectrum of the $K^0 p$ subsystem at 1530 ± 5 MeV/c² with a significance of 4 – 6 standard deviations, depending on background assumptions. The upper limit of 18 ± 4 MeV/c² (FWHM) for its width is given by the experimental resolution. The corresponding total cross section is estimated to be about 0.4 ± 0.1(stat) ± 0.1(syst) μb. Since a resonance in this subsystem must have strangeness S = + 1 we claim it to be the $\Theta^+$ state for which very recently evidence was found in various experiments.

_______________________________________________________________________

[1] Corresponding author: W. Eyrich, Physikalisches Institut, Friedrich-Alexander-Universität Erlangen-Nürnberg, Erwin-Rommel-Str. 1, 91058 Erlangen, eyrich@physik.uni-erlangen.de

[2] Present address: Department of Physics, Kyoto University, Japan

[*] Supported by German BMBF and FZ Jülich

**Introduction**

As presently understood, QCD does not forbid the existence of states other than quark-antiquark and three-quark systems as long as they form colour singlets. A priori, there should be no reason for a strong suppression of exotic states. In fact already in the early phase of QCD motivated models systems consisting of more than three quarks, in particular five quark systems have been discussed [1]. In more recent theoretical publications the possible existence of such pentaquark states has been worked out based on specific assumptions and production scenarios also including heavy quarks (e.g. [2, 3] and others). One of the most cited publications by Diakonov, Petrov and Polyakov [4] is based on the soliton model assuming an antidecuplet as third rotational excitation in a three flavour system. The corners of this antidecuplet are occupied by exotic pentaquark states with the lightest state having a mass of $\approx 1530$ MeV/c$^2$, strangeness +1, spin 1/2 and isospin 0. This state, originally known as the Z$^+$, has more recently been renamed $\Theta^+$. In this model the mass of the $\Theta^+$ is fixed by the N$^*$ resonance at 1710 MeV/c$^2$, which is assumed to be a member of the antidecuplet. The most striking property of the $\Theta^+$ resonance is the predicted narrow width of $\Gamma < 15$ MeV/c$^2$, which according to ref. [4] is connected with a narrow width of the 1710 MeV/c$^2$ N$^*$ resonance of 50 MeV or less. With the predicted quark content for the $\Theta^+$ of $uudd\bar{s}$ this pentaquark resonance is expected to decay into the channels $K^+n$ and $K^0p$.

The first report on the discovery of a narrow resonance in the expected mass region came from the LEPS collaboration at SPring8 [5] where in the $\gamma K^-$ missing mass spectrum of the reaction $\gamma n \rightarrow K^+K^-n$ on $^{12}C$ a narrow resonance was observed at $1.54 \pm 0.01$ GeV/c$^2$ with a significance of 4.6 $\sigma$ and an upper limit for the width of $\Gamma = 25$ MeV/c$^2$. This was confirmed by the DIANA collaboration at ITEP [6] which observed a narrow resonance with a mass of $1539 \pm 2$ MeV/c$^2$ and a width of $\sigma = 3$ MeV/c$^2$ in the $K^0p$ invariant mass spectrum in reanalysed data from the reaction $K^+Xe \rightarrow K^0p\,Xe^{'}$ with a quoted evidence of 4.4 $\sigma$.

In the meantime several other experiments have presented successful observations in the mass region between 1526 and 1555 MeV/c$^2$. The CLAS collaboration [7] reported on a narrow peak in the $K^+n$ system produced in the reaction $\gamma d \rightarrow pnK^+K^-$ at a mass of $1542 \pm 2$ MeV/c$^2$ and a width of $\Gamma < 21$ MeV/c$^2$ and a narrow peak around 1555 MeV/c$^2$ from the reaction $\gamma p \rightarrow nK^+K^-\pi^+$ [8]. A further $\gamma$-induced measurement of the reaction $\gamma p \rightarrow nK^0K^+$ has been reanalysed by the SAPHIR collaboration [9]. The observed mass of $1540 \pm 4 \pm 2$ MeV/c$^2$ and the width of $\Gamma < 25$ MeV/c$^2$ is in agreement with the recent experiments.

Moreover evidence comes from neutrino-scattering on nuclei investigating the $pK_S^0$ system at ITEP [10], where a peak was found at 1533 MeV/c$^2$. Very recently positive results were also reported from the HERMES collaboration [11], where a narrow baryon state was found at 1528 MeV/c$^2$ in quasi-real photo production on a deuterium target in the decay channel $pK_S^0 \to p\boldsymbol{p}^+\boldsymbol{p}^-$, and from the SVD collaboration [12], where in the $pA$ interaction a peak shows up at 1526 MeV/c$^2$ in the $K_S^0 p$ system.

In this paper we report on the search for the $\Theta^+$ resonance using the COSY-TOF experiment. Within the framework of the hyperon production program at COSY-TOF [13, 14] the reaction $pp \to \Sigma^+ K^0 p$ has been measured exclusively. Data were taken at several beam momenta ($p_{\text{beam}}$ = 2.85, 2.95, 3.20 and 3.30 GeV/c) covering the full phase-space. From the beginning of these measurements in 1998 these investigations were partly motivated by the predictions of Diakonov, Petrov and Polyakov [4] of a narrow exotic pentaquark state around 1530 MeV/c$^2$, which should show up in the invariant mass spectrum of the $K^0 p$ subsystem. After an upgrade of the apparatus and significant developments of the COSY beam a detailed search for the $\Theta^+$ became possible by obtaining reasonable event samples for the reaction $pp \to \Sigma^+ K^0 p$, which allowed detailed analysis of the subsystems. Data were taken predominantly at a beam momentum of $p_{\text{beam}}$ = 2.95 GeV/c, corresponding to an excess energy of 126 MeV. This limits the invariant mass spectrum of the $K^0 p$ -system between the threshold value of 1436 MeV/c$^2$ and an upper value of about 1562 MeV/c$^2$. Accordingly an optimal ratio between a possible resonance signal around 1530 MeV/c$^2$ and the non resonant background is expected [15]. A deviation from a smooth invariant mass spectrum of the $K^0 p$ system was already observed in a first measurement performed in 2000, but the extracted event sample was too small for a definitive statement [14]. To improve the statistical significance a second production run at the same beam momentum was performed in 2002, providing in addition the possibility of searching for systematic deviations between both independent measurements.

For the beam momenta of 3.2 GeV/c and 3.3 GeV/c which have been measured in addition the obtained event samples are too small to search for a resonance on a strong non resonant background. We want to mention that data were also taken of the reaction channel $pp \to \Sigma^+ K^+ n$. This channel is in principle also suited for the $\theta^+$ search. However, mainly due to the limited efficiency of the neutron detector used in this case the event samples are also too small for a detailed analysis of the spectra of interest.

**Experimental setup and analysis**

In the production runs in 2000 and 2002 reported in this paper the time-of-flight spectrometer COSY-TOF was used in its 3 m version [16]. The reaction $pp \rightarrow \Sigma^+ K^0 p$ is induced by focussing the extracted proton beam on a spot of about 1 mm $\varnothing$ on a liquid hydrogen target with a length of 4 mm [17]. The geometrical reconstruction of the related tracks and vertices is mainly realized by the start-detector system, a scheme of which is shown in Fig.1 together with an event of the type $pp \rightarrow \Sigma^+ K^0 p$ with a subsequent decay of the $K^0$ as a $K_S$ into a $\boldsymbol{p}^+ \boldsymbol{p}^-$ pair and the delayed decay of the $\Sigma^+$ into a $n\boldsymbol{p}^+$ pair.

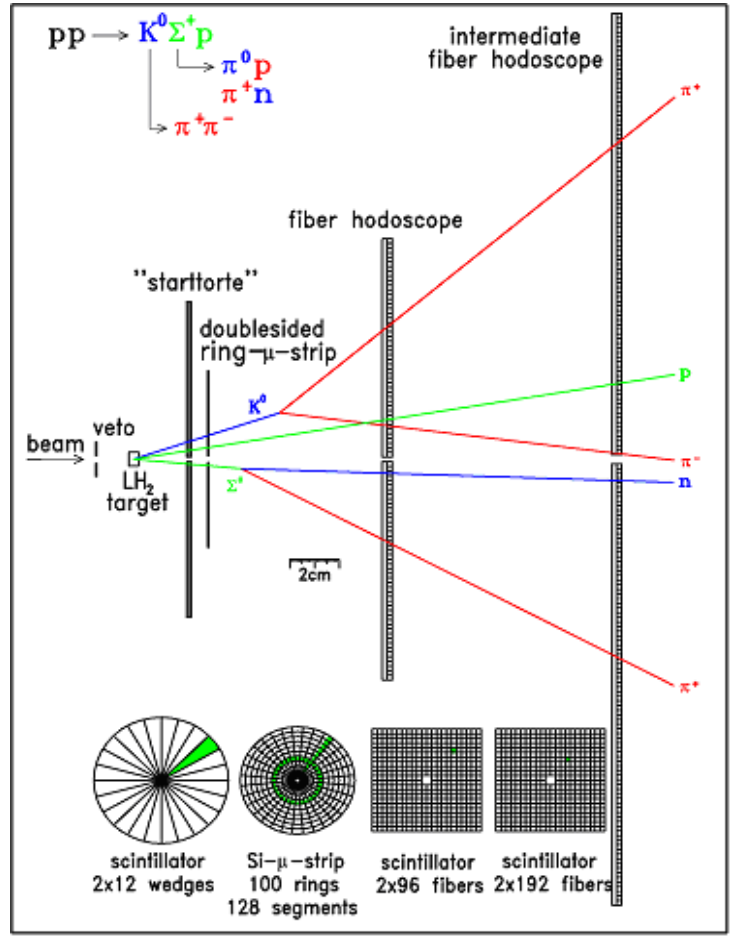

Figure 1: Scheme of the Start detector system together with an event of the reaction $pp \rightarrow \Sigma^+ K^0 p$.

The events of interest are identified by these delayed decays. The reconstruction of the $K_S^0$ and its decay vertex occurs via the tracks of its decay products $\boldsymbol{p}^+ \boldsymbol{p}^-$ by two scintillating fibre hodoscopes. The decay kinematics and angular distributions allow a clear separation from the remaining background, which is dominated by the reaction $pp \rightarrow K^+ \Lambda p$. The $\Sigma^+$ hyperon together with its delayed decay into $\boldsymbol{p}N$ is characterized by a track with a clear kink and is detected via a double sided silicon microstrip detector close to the target. The momenta of the reconstructed particles are calculated directly from the extracted directions ("geometry spectrometer") using momentum and energy conservation. Since there are usually several possible geometrical combinations and hence kinematical solutions for each event, a missing mass analysis is applied for both the mass of the $\Sigma^+$ using the tracks of the primary reaction products and the mass of the $K_S^0$ determined by using the information of the tracks of its decay products. This contains two overconstraints. To find the best solution, both masses are required to be best-fitted simultaneously. Events outside the phase space of the reactions of interest were rejected. Geometrical cuts on the tracks and decay vertices were used to suppress the background. By varying these cuts and performing Monte Carlo simulations in parallel it was carefully checked that the restrictions used do not influence the results concerning the observables of the reaction of interest.

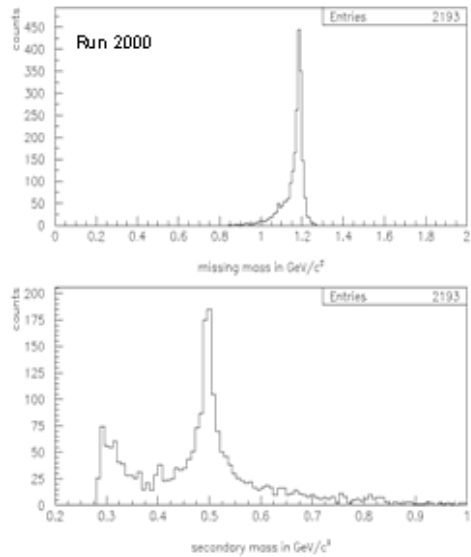

Figure 2: Reconstructed mass of $\Sigma^+$ (upper part) and $K_S^0$ (lower part) obtained from the 2000 run.

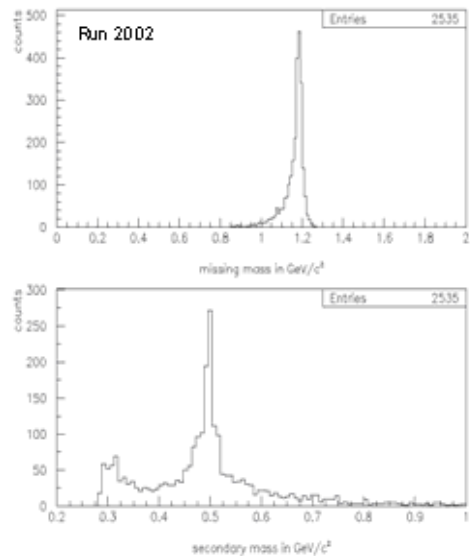

Figure 3: Reconstructed mass of $\Sigma^+$ (upper part) and $K_S^0$ (lower part) obtained from the 2002 run.

Both measurements show clear mass distributions peaking at the related corresponding masses of the $\Sigma^+$ ("primary mass") and $K^0$ ("secondary mass"), respectively, and they are identical within the statistical fluctuations as demonstrated in Fig. 2 and Fig. 3.

To get very clean samples for further investigations of the reaction of interest cuts on the resulting mass peaks have been applied. This is demonstrated in Fig. 4 where the spectra of the two runs are summed up. The upper and lower part show the sum of the spectra of Fig. 2 and Fig. 3. Applying the indicated cut (dashed lines) on the $K^0$ mass (actually between 480 MeV/c$^2$ and 510 MeV/c$^2$), a significant reduction of the width of the $\Sigma^+$ peak is obtained as can be seen in the middle part of Fig. 4 compared to the upper part. The same holds the other way round by cutting on the $\Sigma^+$ peak (dashed lines). Finally the cuts shown on the $K^0$ mass and the $\Sigma^+$ mass lead to two samples of 421 and 518 events for the two runs respectively, and accordingly 939 events for the total sample which is used for further analyses.

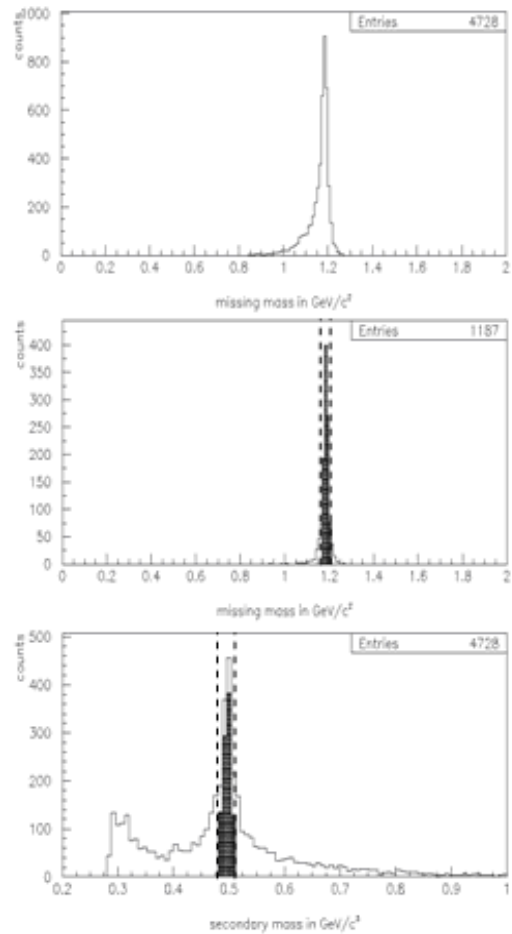

Figure 4: Reconstructed masses of $\Sigma^+$ and $K^0_S$ for the summed data from the run 2000 and the run 2002, respectively. The upper and lower part show the sum of the spectra of Fig. 2 and Fig. 3. The $\Sigma^+$ spectrum in the middle part corresponds to the indicated cut on the $K^0$ mass. Further explanation see text.

Extensive Monte Carlo simulations were performed to control and to optimize the analysis chain and the cuts shown above. Moreover they were used to deduce the resolution in the various observables. The resolution of the $\Sigma^+$ mass and the $K^0$ mass of the simulated data is in quantitative agreement with the real data. For the $K^0\,p$ invariant mass which is relevant for the search for a possible narrow pentaquark state an overall mass resolution of $18 \pm 3$ MeV/c$^2$ (FWHM) has been deduced from Monte Carlo simulations. This resolution is similar to most of the experiments which reported up to now on a narrow state in the *KN* system which might be interpreted as a pentaquark state.

**Results**

To search for a possible resonance the data of the two runs at a beam momentum of $p_{beam} =$ 2.95 GeV/c have been investigated both separately and in sum. In Fig. 5 the invariant mass spectra of the $K^0\,p$ system are shown. They cover the full kinematical range corresponding to the excitation energy of 126 MeV. The shape of all three spectra is very similar. Within statistical fluctuations the spectra from the 2000 (top figure) and 2002 (middle) runs are identical.

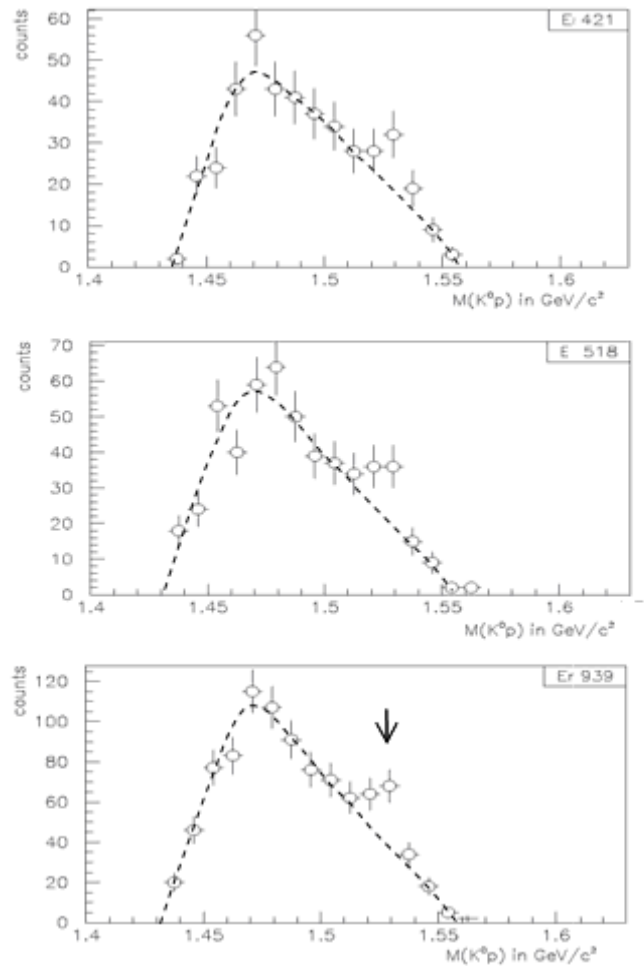

Figure 5: Invariant mass spectrum of the $K^0p$ subsystem obtained from the 2000 data (upper part), the data from 2002 (middle part) and the sum of both (lower part) together with a fitted background.

There is an obvious deviation from a smooth distribution in the spectra of both runs and in the spectrum of the summed samples (bottom) around 1.53 GeV/c$^2$ (indicated by the arrow in the summed spectrum).

Assuming a smooth background as obtained by a polynomial fit excluding the region between 1.51 GeV/c$^2$ and 1.54 GeV/c$^2$ (dashed curves in Fig. 5) the significance of the signal can be deduced. Following roughly the argumentation given in Ref. [11] three different expressions for the significance of the peak in the summed spectrum (Fig. 5 bottom) have been considered. The first alternative is the naïve estimation $N_S / \sqrt{N_B}$ where $N_S$ is the number of events corresponding to the signal on top of the fitted background and $N_B$ is the number of events corresponding to the background in the chosen area. This method was used in Refs [5, 6, 7, 9, 10]. In the present case this leads to a significance of 5.9 σ on the basis of an interval of ± 1.5 σ around the peak value of 1530 MeV/c$^2$. This estimator however neglects the statistical uncertainty of the background and therefore usually overestimates the significance of the peak [11, 18]. A more conservative method which is reliable for cases where the background is smooth and well fixed in its shape uses the estimator $N_S / \sqrt{N_S + N_B}$. In our case this method leads to a significance of 4.7 σ. The third expression taking into account the full uncertainty of a statistically independent background which should underestimate the significance of the signal is given by $N_S / \sqrt{(N_S + N_B) + N_B}$. This leads to a value of 3.7 σ.

Because the measurement presented and the event sample extracted from it cover the full phase space of the reaction products, an investigation of the corresponding Dalitz plot is possible. This facilitates the search for artefacts which may fake a signal. In Fig. 6 the Dalitz plot based on the 939 events corresponding to the summed spectrum of Fig. 5 is shown. The peak around 1.53 GeV/c$^2$ identified in the $K^0 p$ invariant mass spectrum should show up in the ideal case as a band in the Dalitz plot at the corresponding squared mass around 2.34 GeV$^2$/c$^4$ as indicated by the arrows in both distributions. As expected due to the low number of events there is only a slight indication for a band. But more importantly in both distributions there is no indication of an artefact which could give rise to a faked signal in the $K^0 p$ mass spectrum. Especially there is no indication of a possible final state interaction in the crossed channel $\Sigma^+ p$ which in principle could influence the strength in the interesting region of the $K^0 p$ system. Moreover the strength concentrated below a squared mass of about 2.2 GeV$^2$/c$^4$ in the $K^0 p$ channel shows no influence on the region of interest.

It should also be recognized that according to the low excess energy of 126 MeV the influence of a possible excitation of $\Sigma^*$-resonances is excluded in our case.

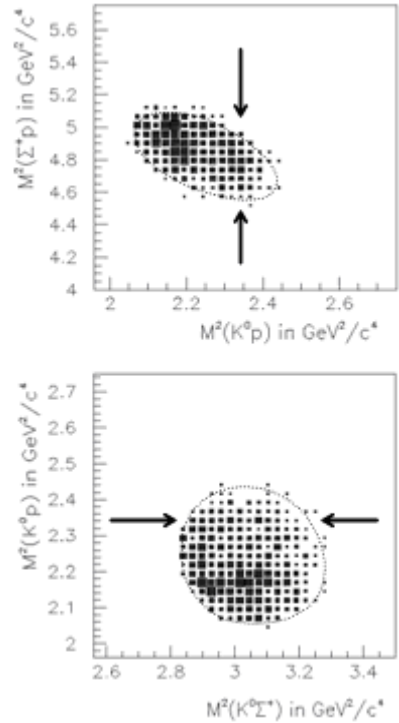

Figure 6: Dalitz plots for the full sample at a beam momentum of 2.95 GeV/c. The dotted lines show the phase space limits. The indicated arrows correspond to an invariant mass for the $K^0p$ system of 1.53 MeV/c$^2$.

To correct for the efficiency of the detector and the analysis, Monte Carlo simulations were used. The correction function is very smooth giving some enhancement at the edges of the phase space. In Fig. 7 the efficiency-corrected $K^0p$ invariant mass spectrum corresponding to the total sample is shown. In comparison to the uncorrected spectrum shown in Fig. 5 there is no major difference. Again there is a significant peak around 1.53 GeV/c$^2$ on top of a smooth background. For a more quantitative analysis a polynomial fit on the background and a Gaussian for the remaining signal are used (Fig.7 dotted lines). This yields a peak value of $1530 \pm 5$ MeV/c$^2$. The deduced width of $18 \pm 4$ MeV/c$^2$ (FWHM) is in agreement with the value of the Monte Carlo analysis and accordingly only an upper limit for the physical width of the observed peak can be quoted.

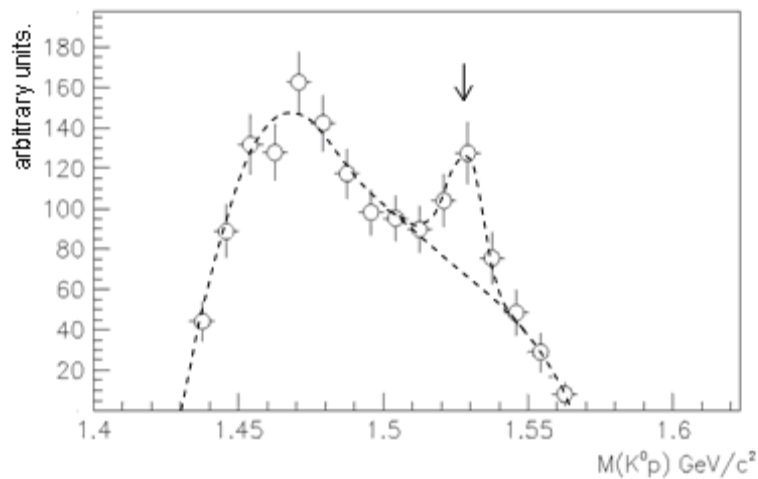

Figure 7: Efficiency corrected invariant mass spectrum of the $K^0p$ subsystem obtained from the full sample.

The cross section of the observed peak around 1530 MeV/c$^2$ has been estimated by comparing with the measured total cross section of the reaction. The normalisation was deduced by comparison with the elastic $pp$ scattering which was measured simultaneously and for which precise data are available from the COSY-EDDA experiment [19]. For the observed peak at 1530 MeV/c$^2$ we deduce a cross section of $0.4 \pm 0.1$ (stat.) $\pm 0.1$ (sys.) *mb* . This value is in rough agreement with theoretical estimations by Polyakov *et al.* [15] and Liu and Ko [20], where a total cross section in the order of 0.1 - 1 *mb* is predicted for the $\Theta^+$ production in the threshold region in $pp$ and $pn$ induced reactions.

**Summary and Outlook**

The COSY-TOF experiment provides evidence for a narrow resonance in the $K^0 p$ system at a mass of $1530 \pm 5$ MeV/c$^2$ from the exclusively measured reaction $pp \to \Sigma^+ K^0 p$ . The extracted width of about 18 MeV/c$^2$ reflects the experimental resolution. Since a resonance in this subsystem must have strangeness S = + 1 we claim it to be the $\Theta^+$ state. The significance of the signal seen in the invariant mass spectrum of the $K^0 p$ subsystem was deduced to be between 3.7 $\sigma$ and 5.9 $\sigma$, depending on the background estimation. This is the first evidence on the $\Theta^+$ resonance from an elementary hadron hadron induced reaction.

To get more insight into the nature of the $\Theta^+$ resonance in an upcoming run the COSY-TOF experiment will use a polarized beam and in a further step also a deuterium target to investigate the reaction channel $pn \to \Lambda K^0 p$ for which our apparatus should be optimally suited. Moreover plans exist to use a polarized beam in combination with a polarized target to deduce the parity of this resonance [21].

**Acknowledgement**

We want to thank very much the COSY accelerator team for the preparation of the excellent proton beam and the good cooperation during the beam time. We are grateful to M. Polyakov for many fruitful discussions. This work is based in part on the doctoral thesis of Marcus Wagner. We gratefully acknowledge support from the German BMBF and the FZ Jülich.